\documentclass[journal]{vgtc}                     

\onlineid{1196}

\graphicspath{{figs/}{figures/}{pictures/}{images/}{./}} 

\usepackage{microtype}                 
\PassOptionsToPackage{warn}{textcomp}  
\usepackage{textcomp}                  
\usepackage{mathptmx}                  
\usepackage{times}                     
\usepackage{cite}                      
\usepackage{tabu}                      
\usepackage{booktabs}                  
\usepackage{soul}
\usepackage{comment}
\usepackage[dvipsnames,svgnames]{xcolor}
\usepackage{academicons}
\usepackage{tikz} 
\definecolor{lime}{HTML}{A6CE39}
\usepackage{wasysym}
\usepackage{multirow}
\usepackage{svg}
\usepackage{expdlist}
\usepackage{marvosym}
\usepackage{pdfpages}
\usepackage{amssymb}
\usepackage{flushend}

\usepackage{xspace}
\newcommand{\ie}{i.e.,\@\xspace}
\newcommand{\eg}{e.g.,\@\xspace}

\newcommand{\inlinevis}[3]{\raisebox{#1}[0pt][0pt]{\includegraphics[height=#2]{#3}}}

\newlength{\picturewidth}

\DeclareUnicodeCharacter{0301}{\'{e}}

\def\nbFiguresRelevant {808 }
\def\nbFiguresGenderCollected {2,162 }
\def\nbScientificFiguresGenderCollected {1,962 }

\def\KrippendorffGen{0.6287}
\def\nbEncodings{577}
\vgtccategory{Research}

\vgtcpapertype{Theoretical \& Empirical}

\title{Eleven Years of Gender Data Visualization:\texorpdfstring{\\}{ } A Step Towards More Inclusive Gender Representation}

\author{\authororcid{Florent Cabric}{0000-0002-9326-9441}, \authororcid{Margrét Vilborg Bjarnadóttir}{0000-0003-2955-1992}, \authororcid{Meng Ling}{0000-0001-6597-5448}, \authororcid{Guðbjörg Linda Rafnsdóttir}{0000-0003-2662-5773}, and \authororcid{Petra Isenberg}{0000-0002-2948-6417}}
\authorfooter{
\item
 Florent Cabric and Petra Isenberg are with Université Paris-Saclay, CNRS, Inria, LISN, France. E-mail: \{florent.cabric\textbar petra.isenberg\}@inria.fr.
 \item
 Margrét Vilborg Bjarnadóttir is with the Robert H. Smith School of Business, University of Maryland, College Park, USA. E-mail: mbjarnad@umd.edu.
\item
Meng Ling is with the Ohio State University, USA. E-mail: ling.253@osu.edu.
\item
Guðbjörg Linda Rafnsdóttir is with the Faculty of Social and Human Science, University of Iceland, Reykjavik, Iceland. E-mail:glr@hi.is.
}
\shortauthortitle{Cabric \MakeLowercase{\textit{et al.}}: Gender Visualization}

\abstract{%
We present an analysis of the representation of gender as a data dimension in data visualizations and propose a set of considerations around visual variables and annotations for gender-related data. Gender is a common demographic dimension of data collected from study or survey participants, passengers, or customers, as well as across academic studies, especially in certain disciplines like sociology. Our work contributes to multiple ongoing discussions on the ethical implications of data visualizations. By choosing specific data, visual variables, and text labels, visualization designers may, inadvertently or not, perpetuate stereotypes and biases. Here, our goal is to start an evolving discussion on how to represent data on gender in data visualizations and raise awareness of the subtleties of choosing visual variables and words in gender visualizations. In order to ground this discussion, we collected and coded gender visualizations and their captions from five different scientific communities (Biology, Politics, Social Studies, Visualisation, and Human-Computer Interaction), in addition to images from Tableau Public and the Information Is Beautiful awards showcase. Overall we found that representation types are community-specific, color hue is the dominant visual channel for gender data, and nonconforming gender is under-represented. We end our paper with a discussion of considerations for gender visualization derived from our coding and the literature and recommendations for large data collection bodies. A free copy of this paper and all supplemental materials are available at \url{https://osf.io/v9ams/}.
}


\teaser{
  \centering
  \includegraphics[width=0.9\linewidth]{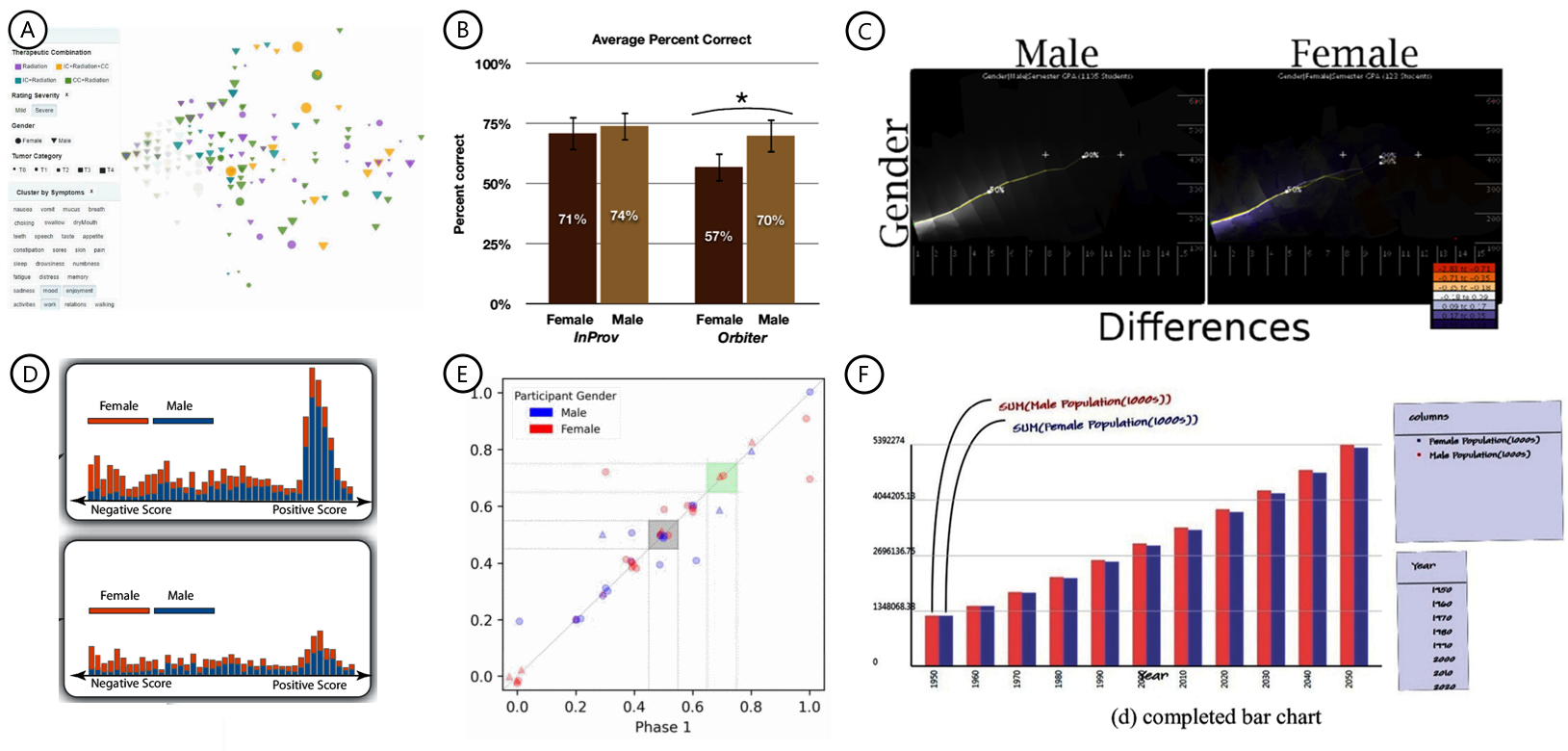}
  \vspace{-1em}
  \caption{Sample of uncommon (top) and common (bottom) encodings to represent gender with (A) a scatter plot using shapes to represent gender \cite{Floricel:2021}, (B) a bar chart with dark and light brown colors \cite{Borkin:2013}, (C) separated charts (heatmaps) \cite{Trimm:2012} and (D, E, and F) bar charts and scatterplot using pink/red colors to represent women and blue to represent men \cite{Wall:2021,Walny:2012,Wu:2012}. Image permissions: \textcopyright\ IEEE}
  \vspace{-0.5em}
  \label{fig:teaser}
}
\begin{document}


\firstsection{Introduction}

\maketitle

Gender is not only a personal topic but also an important element of political and social debate. As many strive towards more inclusive organizations and a more inclusive society in general, the communication of gendered topics and differences, including the collection, communication, and use of gender in data, is more important than ever.
For the visualization community, gender is important because it is a common dimension in datasets about people. Visualization designers are, therefore, regularly challenged with finding appropriate visual channels and language for gender representation. The appropriate representation of marginalized groups (such as non-binary people) is also a particularly important consideration in drawing attention to the specific problems these groups may face.
We apply scientific approaches to review the current state of gender visualization to raise awareness and evolve the discussion of how to represent data on gender.

Data visualizations are often seen as objective representations of reality and are sometimes the only way to understand a phenomenon. Yet, representation styles are not neutral. A chosen mapping between visual channels and data can be the source of errors or unexpected behavior \cite{Szafir:2018}. For example, colors can perpetuate stereotypes such as ``pink for girls and blue for boys'' \cite[Chapter 4]{Dignazio:2020}, influence decision-making processes \cite{Kliger:2012}, and have strong emotional associations \cite{Demir:2020} (for example, ``red'' is associated with anger for English speakers \cite{Fugate:2019}). Moreover, research has provided extensive evidence that color perception \cite{Tilburg:2015}, association \cite{Jonauskaite:2019}, and preference \cite{FortmannRoe:2013,Jonauskaite:2019,Lobue:2011} are socially constructed and transform what could otherwise be an individual choice into standardized behavior. These stereotypes can appear very young and significantly impact children's development \cite{Karniol:2011}.

It is, therefore, important for the visualization community to engage in continuing discussions on how to represent data dimensions that we are morally obligated to treat carefully.
To ground our discussions, we present an 11-year review of how certain scientific communities visually represent gender in their papers, and we contrast our analysis with studies of images extracted from Information Is Beautiful Awards \cite{DVS:2022:IIBA} (IIBA) and Tableau Public \cite{TableauPublic:2022} (TP). Our work focuses on how authors differentiate between different genders in their visualizations and which words they use to describe their gender data.
While exploratory, our analysis targeted specific research questions:
\begin{itemize}
\vspace{-0.0em}
\setlength\itemsep{-0.4em}
\item \textbf{How do scientists visually represent gender in their papers?} This analysis of the gender visual representation can help the scientific community inspect its own practices and lead to a discussion of future approaches to the representation of gender data.
\item \textbf{What words do scientists use to describe gender in their visualizations and captions?} Which genders are talked about and represented in data visualizations?
\item \textbf{What between and within-community patterns can be found?} Scientific communities write papers in their own style, including figures and data representation. Do we see differences between and within communities when comparing the visualizations of gender/sex they produce? 
\item \textbf{Have there been any shifts in gender visualization between and within communities during the last decade?} Gender perception and representation are fast-moving, and the previous ten years have been marked by the growth of movements that ask for a more inclusive society. Have scientists shifted their own practices in response to these ongoing discussions?
\end{itemize}

Our work results from a collaboration between visualization, management science and statistics, and sociology researchers. We contribute to the 
ongoing discussions on the ethics of data visualization \cite{Correll:2019:Ethics}, focusing not only on a socially important 
topic but on a specific data dimension within that topic. We hope to provide initial discussion points that will promote the development of ongoing and potentially changing recommendations and considerations for representing gender data; these points may also influence how we collect and represent data on gender in our studies. In summary, this paper contributes to the following:

\begin{itemize}
\setlength\itemsep{-0.4em}
\item A set of \nbFiguresRelevant scientific figures representing human gender data.
\item An analysis of visual representations, practices, and words used to communicate about different genders in five different scientific communities and two non-scientific communities.
\item A set of considerations for the representation of gender in data visualizations.
\end{itemize}

\section{Background: A Social Sciences Perspective on the Representation of Gender}
Recent studies show that while women’s participation in the workforce and professional education has increased substantially, gender stereotypes remain. Little has changed over the last 30 years when it comes to gender stereotypes regarding traits, role behaviors, occupations, or physical characteristics \cite{Haines:2016:TimesAreChanging}. It is essential to understand what causes and maintains this stagnation, which is concurrent with an increased emphasis on gender equality. As personal dispositions and social attitudes are influenced by the individual’s social context \cite{Haines:2016:TimesAreChanging,haslanger2017gender}, it is vital to understand how visual effects reflect and reproduce our ideas on gender. Even though we do not offer theories in this paper, we point out that according to the cultural lag hypothesis \cite{diekman2010and},
gender attitudes and beliefs are likely to lag behind societal changes.

\subsection{Beyond binary genders}
Despite the increasing cross-disciplinary acceptance of the fact that there are not only two genders and that many people identify themselves outside of this binary
\cite{ainsworth2015sex,winter2016transgender}, much of the world is still defined as more or less in binary terms. Thus, even though increasing numbers of people cannot define themselves within the binary gender structure, it seems difficult in both research and everyday life for people to leave the binary stereotypes behind. Cunningham and Macrae \cite{cunningham2011colour} and Nash and Sidhu \cite{nash2023pink} show that despite legislative attempts to eliminate gender stereotyping from society, the stereotyped gender identities are still robust due to their overuse in everyday life and the propensity to evaluate people on the basis of their binary sexes. For instance, showing children in different colors sets the stage for the automatic activation and expression of gender stereotypes. In other words, visual gender stereotypes perpetuate general stereotypes about gender characteristics. 

Gender is usually understood as a set of socially constructed norms, attitudes, feelings and behaviors attributed in a binary way to men or women. For people who do not fit into this binary representation, digital media and sites for learning have allowed people to overcome isolation, connect, and develop shared languages \cite{Nelson:2022:SocialMedia,Cover:2019:EmergentIdentities,Robards:2020:Tumblr}. In this way, digital media has been a key to individual exploration of identity. Yet, digital media can also bind us in chains of stereotypical identity at the same time as it brings us certain freedoms.

\subsection{Representation of gender in language}

Westbrook and Saperstein \cite{westbrook2015new} show that individual researchers and, increasingly, computers (also those running generative AI models) are often expected to supply a pronoun based on what is known about a given individual. When surveys ask about topics unrelated to gender, respondents are constantly gendered through this binary pronoun fills, constraining responses and reinforces traditional categories. Gendering through pronouns reproduces an understanding that sex and gender are binary and the world is composed only of ``hes'' and ``shes'' \cite{Aikhenvald:2016:HowGenderShapesTheWorld,westbrook2015new}. Even though some scholars have changed their survey measurement methods in response to criticism of the binary approach to gender, there have been fewer changes in national surveys and big databases. 

The gender-capturing categories offered on surveys and official documents have been taken for granted and assumed to reflect natural distinctions between women and men. However, in line with the recent questioning of this binary, the most recent studies make efforts to reflect the diversity of gendered lives by incorporating nonconforming gender identities into the research \cite{vivienne2022social} and by improving the measurement of sex and gender \cite{westbrook2015new} beyond traditional binary options.
The most common approach to collecting respondents' gender is to ask them to identify themselves using a list of gender options, including a write-in question \cite{Bauer:transgender:2017}. However, researchers are still developing the language to reflect this conceptualization accurately.

As these findings indicate, we are in the midst of an active, ongoing discussion on the use and definition of \textit{sex} or \textit{gender} \cite{Hyde:2019}. At the risk of over-generalizing, we can broadly state that health and biological sciences use \textit{sex} in a way restricted to biological purposes, while those in fields such as sociology use \textit{gender} to refer to sociocultural differences. Because the terms \textit{gender} and \textit{sex} are so intertwined, some researchers have proposed abandoning the distinction between them \cite{Muehlenhard:2011:DistinguishingSexGender}. In this paper, for the sake of simplicity of presentation, we use the term \textit{gender} to refer to both sex assigned at birth and how people choose to self-describe. This is in keeping with our approach, as we consider the representation of both sex assigned at birth and self-described gender to be relevant for the visualization and broader scientific community.

\section{Related Work in Visualization}
We see our work as part of ongoing efforts in the visualization community to raise awareness of the ethical dimensions of working with and representing data. This conversation has taken place in research papers \cite{Baumer:2022:OfCourseItsPolitical,Correll:2019:Ethics,Dignazio:2020,Morais:2022:Anthropographics,Peck:2019:DataIsPersonal}, the Vis4Good workshops (\href{https://vis4good.github.io/}{https://vis4good.github.io/}), and in discussions of diversity within the community itself \cite{Rogowitz:2019:MTM,Sarvghad:2022:GenderInVis,tovanich:2022:gender}. Correll \cite{Correll:2019:Ethics} summarizes several important arguments against the neutrality of visualizations: data itself is not neutral, the absence or presence of information is often a deliberate choice, and the choice of visual encoding techniques profoundly impacts what people see or learn from a visualization. Similar concerns have been raised in work on Critical InfoVis \cite{Doerk:2013:CriticalInfoVis}, Visualization Mirages \cite{McNutt:2020:Mirages}, Anthropographics \cite{Morais:2022:Anthropographics}, and work in a political context \cite{Baumer:2022:OfCourseItsPolitical}. To overcome some of these challenges, D\"ork et al.\ \cite{Doerk:2013:CriticalInfoVis} outlined a process to examine visualizations for their subjective and interpretive content critically. The authors use the idea of plurality to examine which perspectives a visualization offers in exploring a topic. For example, gender representations could be studied through this lens of plurality in terms of whether and how nonconforming gender are included and how to explore these individuals' data, especially if the numbers are very low. 

Gender data is inherently about people, so visualizations designers should be careful, as Correll points out \cite{Correll:2019:Ethics}, not to create charts that may be ``cruel'' or ``inhumane'' \cite{Dragga:2001:CruelPiesInhumanity}. Developed to counter this problem, anthropographics are visualizations that aim to promote prosocial feelings or behavior. In the anthropographics design space, Morais et al.\ \cite{Morais:2022:Anthropographics} outline the difference between individuals' genuine attributes and their synthetic attributes. For example, a survey participant’s gender is not always known, but to make a visualization more relatable, the visualization designer may introduce gender as a synthetic attribute; as the authors point out, this introduces an ethical dilemma. The authors' design space itself is also highly relevant to the visualization of gender. It includes seven dimensions: granularity (or level of aggregation), specificity (to what extent people can be distinguished), coverage (how many people are visualized from the dataset), authenticity (how many synthetic attributes are present), realism in the depiction of people, physicality of the representation, and situatedness. In our gender representation review, we mostly saw representations of low granularity, low specificity, low realism, and low physicality -- as is common for many statistical charts that aim to convey broad patterns rather than information about individuals. Nevertheless, we are still left with our original questions about how to represent this data about people best.

\section{Gender Image Dataset and Review}

To begin our discussions of how to represent gender in data visualizations, we first wanted to gain an overview of current practices. Specifically, we focused on the research domain as a primary source of statistical charts to inform our scientific community about best practices. To deepen this discussion, we also compare our results to trends in online visualizations aimed primarily at the public rather than researchers. These considerations all influenced our source selection, as described in more detail below. Regarding time frame selection, we chose to focus our collection on a timeframe covering the years 2012--2022. This period saw a marked emergence of questions about gender in research and in society in general. Our analysis of these last eleven years allows us to study trends or shifts resulting from the ongoing social movement towards broader gender equity and increased inclusivity.

\subsection{Source selection}
We first extracted figures from 11 years of IEEE Visualization and ACM CHI papers as representative venues of the visualization and Human-Computer Interaction (HCI) communities. These venues publish numerous papers with user studies, so we expected them to be a good source of relevant images. However, we also assumed that communities dedicated to studies related to sex or gender would have developed specific visualization practices. We, therefore, chose to be informed by work in domains where gender is not only demographic metadata but a main variable of studies. We searched for suitable publications in Scimago (\href{https://scimagojr.com/}{scimagojr.com}), a journal aggregator that ranks sources based on information from Scopus. We pre-selected journals categorized as ``Gender Studies'' by Scimago and ranked them in descending order according to their 2021 Scimago Journal Ranking Indicator. For the first 30 journals on the list, we analyzed all the papers published in 2021 and 2022 and included a journal if it met the following criteria:
\begin{enumerate}
\setlength\itemsep{-0.4em}
    \item At least 90\% of papers were accessible to us.
    \item At least one issue has been published each year since 2012.
    \item No guidelines existed on how to represent gender in figures.
    \item Around 10\% of papers included figures representing at least two genders.
\end{enumerate}

Three journals from different fields met the criteria: \textit{Biology of Sex Differences} (BSD) \cite{BSD}, \textit{Sex Roles} (SR) \cite{SR}, and \textit{Politics \& Gender} \cite{PG}.

To be selected, public-facing sources needed to cover the entire period, be accessible, be easily scrapable, be used by different kinds of creators (casuals and practitioners), and offer either a search function or description we could analyze. \textit{Tableau Public} (TP) and \textit{Information Is Beautiful Awards} (IIBA) fulfilled all conditions. Unfortunately, some well-known data visualization collections, such as MASSVIS \cite{Borkin:2013:Memorable} and DATA USA \cite{DATAUSA:2022}, did not fit these criteria. MASSVIS collected visualization until 2013 and does not provide text from which to extract the relevant images. DATA USA provides redundant data visualizations in which gender is often represented in identical ways. A summary of our sources and their domains can be found in Table \ref{tab:sources}. 

\begin{table}[tb]
  \caption{The sources for our analysis, the number of extracted images from each source, and the number of coded images.}
  \label{tab:sources}
  \scriptsize%
	\centering%
 \renewcommand{\arraystretch}{1.5}%
 \setlength{\tabcolsep}{2pt}
  \begin{tabu}{@{}X[3,l,p]|X[0.5,c,m]|X[1.6,l,m]|X[0.7,c,m]|X[0.7,c,m]}
  \toprule
   Source Name & Abbr. &  Domain & \#Extrac- cted & \#Coded \\
  \midrule
	ACM Conference on Human Factors in Computing Systems \cite{CHI} \newline & CHI & Human-Computer Interaction & 33651 & 75 \\
	IEEE Visualization and Visual Analytics Conference \cite{VIS} \newline & VIS & Visualization & 14180   & 43\\
	Biology of Sex Differences Journal \cite{BSD} \newline & BSD & Biology &  1953   & 243 \\
	Sex Roles Journal \cite{SR} & SR & Social and Behavioral Science & 872  & 139 \\
	Politics \& Gender Journal\cite{PG} & PG & Political Science & 421 &  77\\
    \midrule
    Showcase of Information Is Beautiful Awards \cite{DVS:2022:IIBA} & IIBA & Public &  100 &30\\
    Tableau Public \cite{TableauPublic:2022} & TP & Public &  100 &66\\
  \bottomrule
  \end{tabu}%
  \vspace{-1.5em}
\end{table}

\subsection{Image extraction}
For research papers, we extracted images from the HTML version when available; otherwise, we used a CNN-trained algorithm \cite{Chen:2021}, to extract figures and captions from PDFs. In total, we collected 51,077 scientific images and their associated captions. We filtered and kept only the images that included at least one of the following words or variations thereof (such as plurals and possessives): 
\begin{itemize}
\setlength\itemsep{-0.4em}
    \item female, feminine, woman, girl, daughter, mother, grandmother
    \item male, masculine, man, boy, son, father, grandfather
    \item gender, transgender, transidentity, sex
    \item non-binarity, non-binary
    \item demographic, anthropographic
\end{itemize}

\subsection{Image coding}\label{sec:coding} 
\label{sec:exc}
\label{sec:visvar}
For each of the remaining \nbFiguresGenderCollected images that contain one or more of the abovementioned words, six coders (four co-authors of this paper and two hired undergraduate students) coded each image according to a set of criteria using a custom image coding tool\footnote{A short demonstration video is available in the supplemental material}. We developed a video tutorial and a user guide for all coders. Each image was coded by a pair of coders, including at most one undergraduate. 

We established the codes in a pilot-coding process, during which we took online figures of gender visualizations and attempted to describe which encodings were used for gender. Bertin's visual variables inspired our codes about the visual encodings \cite{bertinSemiologyGraphicsDiagrams2011a}, but our final set included some additional codes and some merged or renamed ones. For each image, the coders assessed:

\begin{description}[\compact]
\item[Relevancy:] Coders checked the relevance of images based on two main criteria: images had to represent at least two human genders (\ie not animals) and had to be visualizations of data (as opposed to photographs). Coders could also mark images that were not understandable or had an ``other'' reason for exclusion.
\item[Caption:] We ran an algorithm to tag each word associated with gender in each caption (using our search terms plus abbreviations). Each coder verified whether the terms were correctly identified and marked additional ones that were missed. 
\item[Gender Encoding:] For each image, coders specified how the represented genders were differentiated visually. The codes were: 
\begin{description}[\compact\setlabelstyle{\itshape}]
\vspace{-0.75em}
    \item[Color:] selected precisely by coders with an eyedropper tool. As such, the color code encapsulated color hue, color value and color saturation. 
    \item[Shape:] selected precisely by coders from 9 basic shapes or ``other.'' \inlinevis{-1pt}{1em}{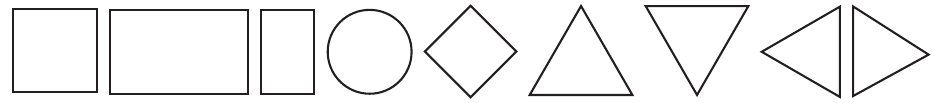}
    \item[Texture:] refers roughly to applying a set of repeated, basic shapes over a given region \cite{hawkins:1970:textural} such as \inlinevis{-1pt}{1em}{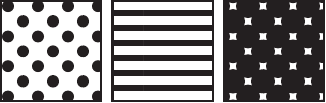}.
    \item[Position:] when gender data is shown in dedicated positions, for example, in dedicated bars in a bar chart.
    \item[Size:] of area mark; for example, \inlinevis{-1pt}{1em}{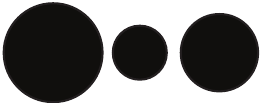}.
    \item[Thickness:] of line marks; for example, \inlinevis{-1pt}{1em}{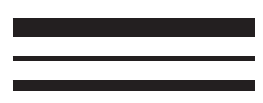}.
    \item[Line Pattern:] applied to lines; for example, \inlinevis{-1pt}{1em}{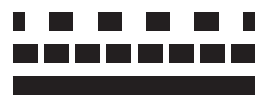}.
    \item[Separated Chart:] when all genders are represented in different charts with their own axes, labels, legends, etc.
    \item[Icons:] more complex shapes frequently used to represent gender, including \Female, \Male, or \FemaleMale.
    \item[Table:] representation of gender as text arranged in tables. 
\end{description}
\item[Visualization Text:] Coders selected words used in the figure itself, such as in legends, axis labels, or annotations. Coders selected these words from a pre-defined list established during our pilot coding or selected ``other'' if a new relevant word was found. When visualizations do not contain text, coders can select ``none.''
\end{description}
\vspace{-0.5em}

Apart from color and shape, all encodings could be marked only as present or absent. For color and shape, coders could specify multiple items, which was useful when images included multiple charts. An image of our coding interface is available in the supplementary material. 

After the first coding phase, our Krippendorff's Alpha (calculated on relevancy and gender encoding) was \KrippendorffGen, which was relatively low. We, therefore, engaged in a recoding phase. If the two coders did not agree, a third expert reviewer coded the image. The author team discussed the few images that none of the coders agreed on and jointly chose the coding.

\section{Results}

Of the \nbScientificFiguresGenderCollected images collected from academic sources, we coded \nbFiguresRelevant as relevant. 
These images belonged to 443 unique papers. To mitigate the problem that single papers could contribute multiple similar-looking images, we removed images from the analysis that had duplicate encoding strategies per paper. As a result, our analysis of academic images is based on a subset of \nbEncodings\ images. 

Of the IIAB and TP images, 96 were relevant (30 \& 66, respectively). As the lack of extracted metadata made tracking the origin of public-facing visualizations difficult, we kept all images for our analysis. We found that only ten of the academic images and seven of the public-facing images represented nonconforming or a ``not reported'' gender. Given this small number of images, it was not possible to provide a meaningful aggregated analysis of the visual representation of nonconforming gender. We will, however, discuss examples in \cref{sec:nonbinaryvis} that we consider particularly relevant. As a result, the following analysis will focus on the representation of binary gender (\ie women and men). We will first study the academic images in detail before contrasting them with the public-facing images.

\subsection{Encodings}\label{sec:encodings}
\begin{figure}[tb]
 \centering
 \includegraphics[width=\columnwidth]{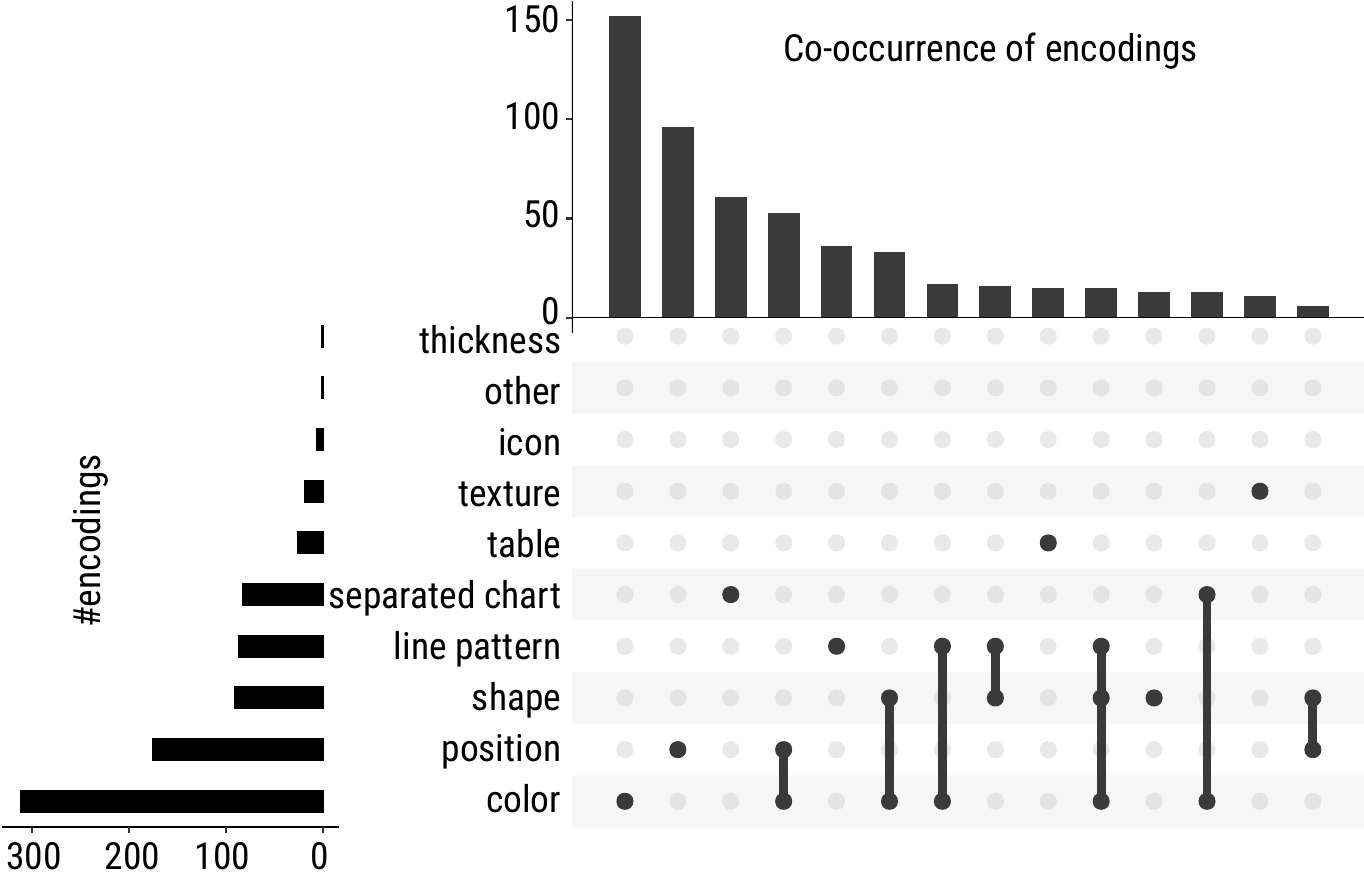}
 \caption{General encodings for visual representations of gender in scientific papers. Minor associations ($<$ 5) have been cropped.}
 \label{fig:encodings_general}
\end{figure}

%

A key focus of our study is how genders were visually distinguished in charts. Of the \nbEncodings\ images from academic papers, 386 distinguished genders with one encoding only (66.9\%), 155 did so with two encodings (26.9\%), and 34 used more than two encodings (5.9\%). When multiple encodings were present, these were not necessarily added to a single visualization, as coders marked 228 images as containing multiple representations. We removed two images that were obviously miscoded (\ie no variable or words were coded for those images). \Cref{fig:encodings_general} summarizes the use of different encodings.

Only 5 of our gender encodings were relatively common. Color encoding was by far the most frequent (in 311 images overall and in 152 images as the only encoding). Position encoding was present in 175 images (alone in 96), especially in bar charts. Shape, line patterns, and the use of separated charts had quite similar counts: 91, 87, and 82 images, respectively (alone in 13, 36, and 61). The remaining encodings (table, texture, icon, other, and thickness) were used 74 times.

The 96 public-facing images in our dataset almost exclusively used color (84\%) to represent and distinguish genders (in 81 images, alone in 65), followed by position (both associated with color (7) and used alone (6)), then icons (9 in total). The use of icons was more common than in scientific images, where icons were used to encode gender in only 6 of \nbEncodings\ images. On the other hand, public-facing images did not use shapes for gender representation.

\subsection{Color}\label{sec:colors}

\setlength{\picturewidth}{.5\columnwidth}
\begin{figure}[tb]
 \centering 
\includegraphics[width=.6\columnwidth]{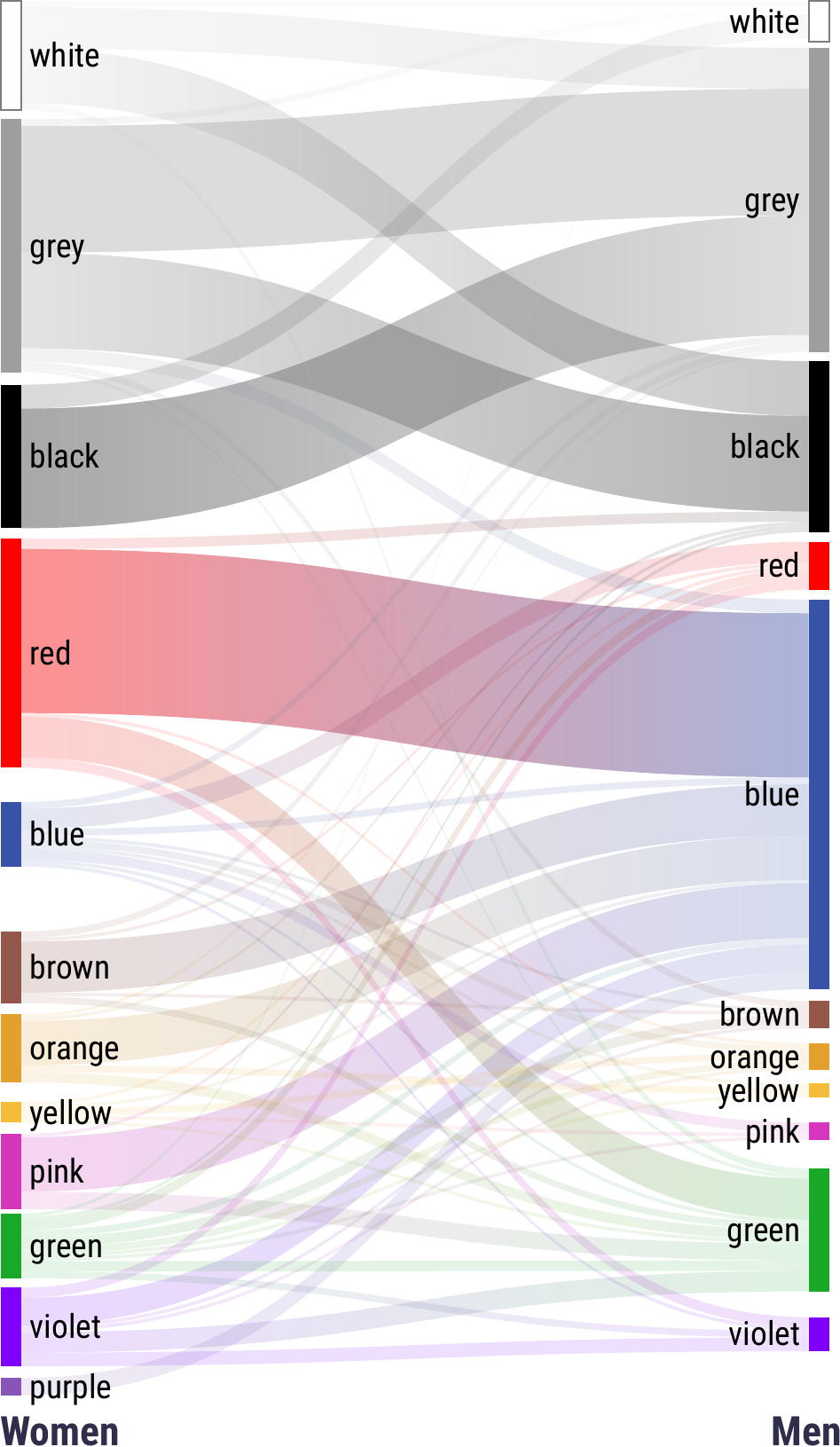}
 \caption{Sankey Diagram with color groups used to represent women (left) and men (right). Ribbons represent the frequency of color combinations present together in the same images.}
 \label{fig:sankey_color}
 \vspace{-1em}
\end{figure}


Color was the main visual variable used to represent gender, which warrants deeper analysis. The coders collected hexadecimal gender colors from each image with an eyedropper. To analyze these colors, we cast them from hexadecimal to the L*a*b* colorspace. Then, each color was named using the ColorNameR package \cite{Beeckman:2021:ColorName}. The package extracted these names from a color glossary of the International Union for the Protection of New Varieties of Plants. It returns for each color a maximum of three pieces of information: intensity, color range, and a single color (\ie the predominant color). For instance, a \textcolor{BlueGreen}{light green blue} is a blue blended with green in light intensity. We collected 354 colors representing women and 350 colors for men. The numbers are not equal because some papers studied different groups of women (\eg pregnant vs. non-pregnant, pre- and post-menopausal) and compared them to a single group of men. We manually renamed some colors returned as green but represented ``perfect'' grey in the RGB colorspace (\ie each RGB component was equal). 

In the following sections, we analyze the color differences by (1) gender, (2) time period, (3) journal type, and (4) targeted audience (scientists or public-facing).

\subsubsection{General analysis}

First, we aggregated colors according to their predominant color (the opaque bars on the left/right of \cref{fig:sankey_color}). The distribution of colors representing women was more balanced than the distribution for men, which seemed concentrated around three color groups: blues, greys, and blacks. Grey was the most common color group for women (21\% of all color encodings), followed by red (19\%). Grey was also common for men (25\%). Blue was the most common color group for men (33\%) but only represented 5\% of all color encodings for women. 

The top of \cref{fig:sankey_color} shows the achromatic color groups. White was more common for women (9\% vs.\ 3\% for men). The achromatic colors were most frequently used together, perhaps for Black and White print. When black was used for men, women were almost exclusively represented with a brighter achromatic color (grey or white). However, when men were represented with grey, women were represented almost equally as often with black or another shade of grey. 

For chromatic colors, one of the most common associations was red for women and blue for men. When a blue color represented men, the color for women were most commonly red but also included pinks, purples/violets, browns, or oranges. In contrast, when women were represented using red colors, men were predominantly either represented in blues or greens.

\subsubsection{Analysis by period}
\begin{figure}[tb]
 \centering 
 \includegraphics[width=\columnwidth]{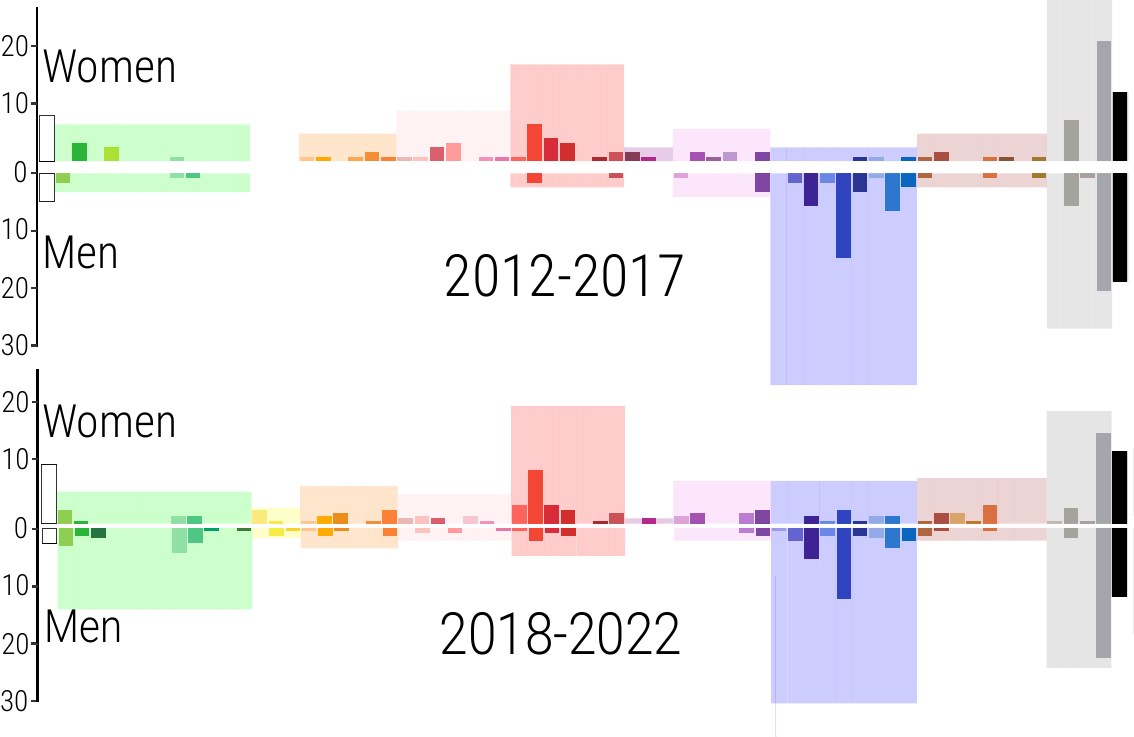}
 \vspace{-1em}
 \caption{Color histogram showing colors used to represent women and men for two time periods:(top) 2012--2017 (bottom) 2018--2022}
 \label{fig:colors_per_year}
  \vspace{-1em}
\end{figure}

To study potential historical trends, we split our dataset into two periods, from 2012--2017 and 2018--2022. The split allowed us to compare earlier to later years and is not based on any specific historic event due to the international authorship of our sample. We grouped colors based on their full names and computed frequency histograms normalized by the number of color codes used in each time period (\cref{fig:colors_per_year}). 

The colors used to represent women remained relatively consistent: for the chromatic colors, red was still dominant for women and blue for men. Green color seemed to have become a little more common to represent men recently, and there seems to be evidence that scientists began to use some orange and pink for men as well. On the other hand, the use of greys became a little less common for both genders. 

\subsubsection{Differences between biology and non-biology journals}

%

We also compared the visualizations from the biology and the non-biology journals in our dataset. We wanted to see if our results might be dominated by the biology journal that not only contributed most images but also often focused on sex rather than gender. 

\begin{figure}[tb]
 \centering 
 \includegraphics[width=\columnwidth]{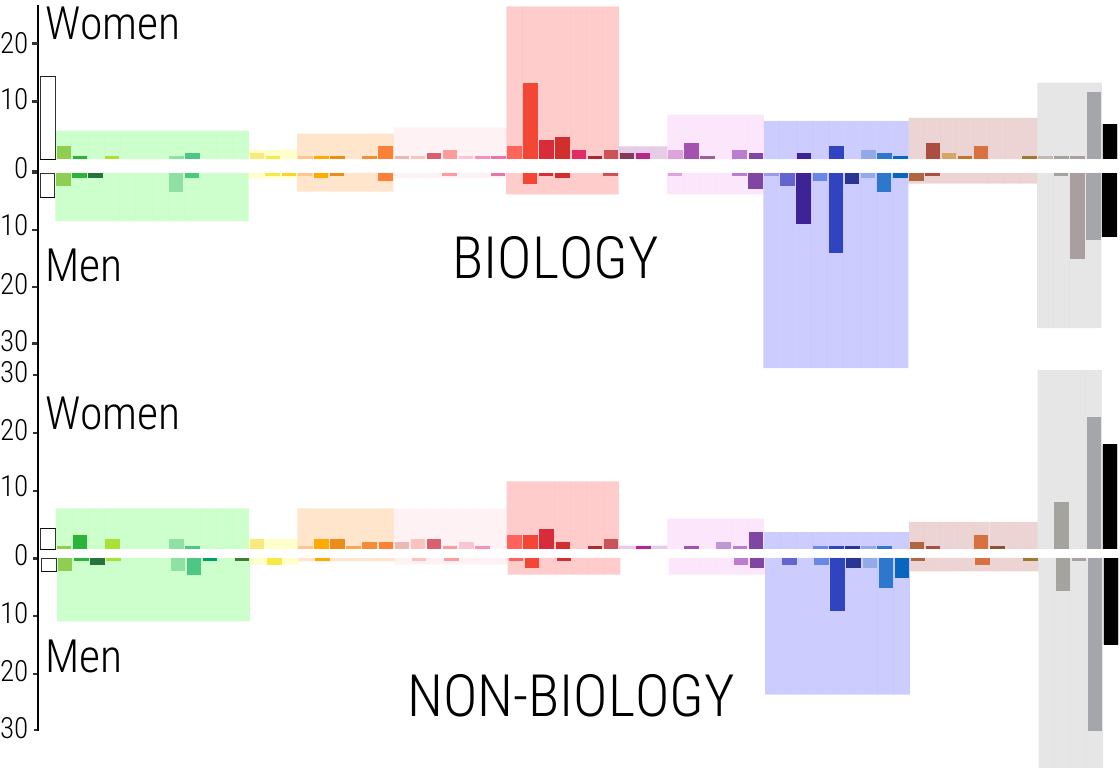}
 \caption{Color histogram showing colors used to represent women and men according to journals: (top) biology, (bottom) non-biology}
 \label{fig:colors_per_journal}
 \vspace{-1em}
\end{figure}

In \cref{fig:colors_per_journal}, we can see that in the biology journal, the predominance of red and blue colors for women and men is quite pronounced; and even higher than the use of any of the achromatic colors. In the non-biology journals, achromatic colors dominated, especially shades of grey and black. We attribute this largely to images from the Politics and Gender and Sex Roles journals. In an analysis of the Sex Roles journal policies, we found that color figures are free of charge for online publication, but they are turned into greyscale figures for printed versions, which might have led authors to favor achromatic images.

Compared to the biology journal, the chromatic colors for women seemed slightly more varied, and greens were relatively more frequent to represent men. Yet, in the non-biology journals, both reds and blues were more dominant for women and men.

%

\subsubsection{Comparison of academic and public-facing images}
\begin{figure}[tb]
 \centering 
 \includegraphics[width=\columnwidth ]{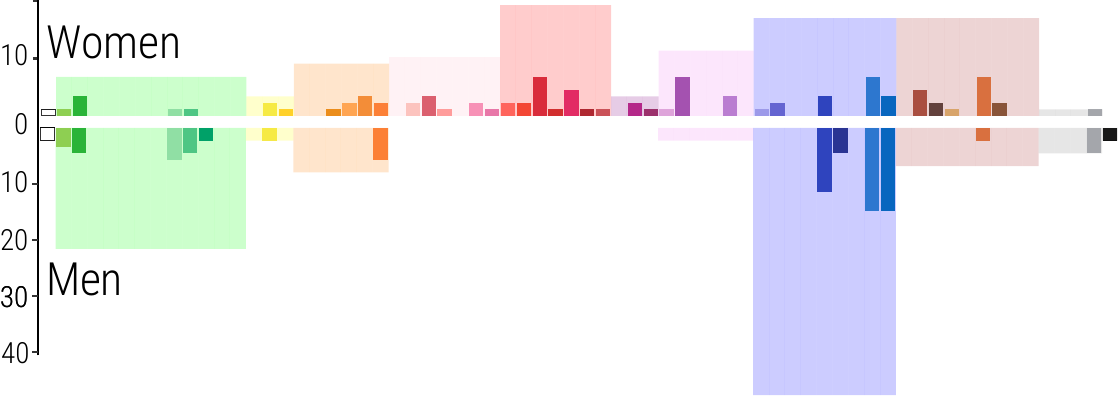}
 \caption{Colors histogram for women (top) and men (bottom) in the public visualizations dataset.}
 \label{fig:colors_public_facing}
\end{figure}

The colors used in public-facing images differed slightly from those used in academic images. Designers of public-facing visualizations rarely used greyscale colors to represent women or men (cf. \cref{fig:colors_public_facing}). The women's color histogram shows a more varied use of chromatic colors than the men one, which is most similar to our observation of the academic figures in the non-biology journals. 
%
\subsection{Shape}
\label{sec:shape}
Shape as an encoding is often used to distinguish single data point categories. For gender, there are some domain-specific shape-related conventions. Kinship diagrams and pedigree charts \cite{Bolin:1999:Perspectives,mcelroy2020signs}, for example, use circles \fullmoon\ for women and triangles $\triangle$ or squares $\square$ for men.
We saw shapes mostly used in scatterplots or line charts. 

We collected 108 shape encodings for women and 104 shape encodings for men. The most common shapes for women were circles \fullmoon\ (38\texttimes) , squares $\square$ (31\texttimes), and diamonds $\diamond$ (30\texttimes). Men were most often represented by squares $\square$ (41\texttimes), circles \fullmoon\ (39\texttimes), and triangles $\triangle$ (23\texttimes). The diamond $\diamond$ shape was much more common for women than men. 

\begin{figure}[tb]
 \centering 
 \includegraphics[width=0.8\columnwidth]{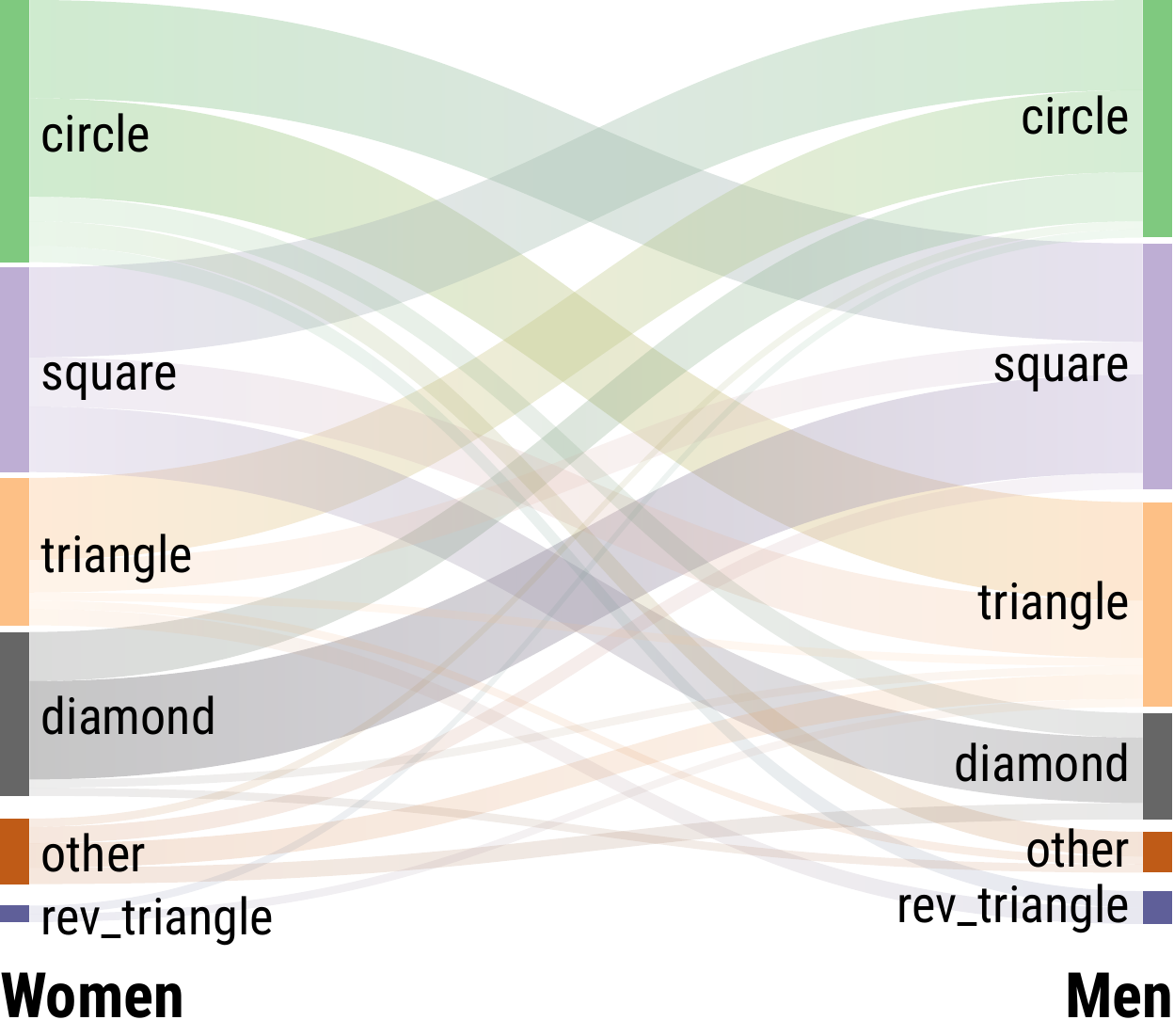}
 \caption{The frequency of different shapes for women (left) and men (right) and their co-occurrence patterns. Rev\_triangle refers to a reversed triangle $\bigtriangledown$. }
 \vspace{-1em}
 \label{fig:shapes_gender}
\end{figure}

The Sankey diagram in \cref{fig:shapes_gender} shows that when women were represented by circles \fullmoon\ the shapes associated with men were the square $\square$ or triangle $\triangle$, mirroring the abovementioned convention. Yet, when men were represented by circles \fullmoon\, we see the same pattern that women were most commonly represented as either triangles $\triangle$ or squares $\square$. 

\setlength{\picturewidth}{.5\columnwidth}
%

As mentioned in Section \ref{sec:encodings}, our dataset of public images did not contain shape representation.

\subsection{Text analysis}
In this section, we examine the most commonly used gender-related words in captions and in the text embedded in images; legends, axis labels, or annotations in particular. We study both the occurrences of individual words on their own and the co-occurrence of several words.

\begin{figure}[tb]
 \centering
 \includegraphics[width=.9\columnwidth]{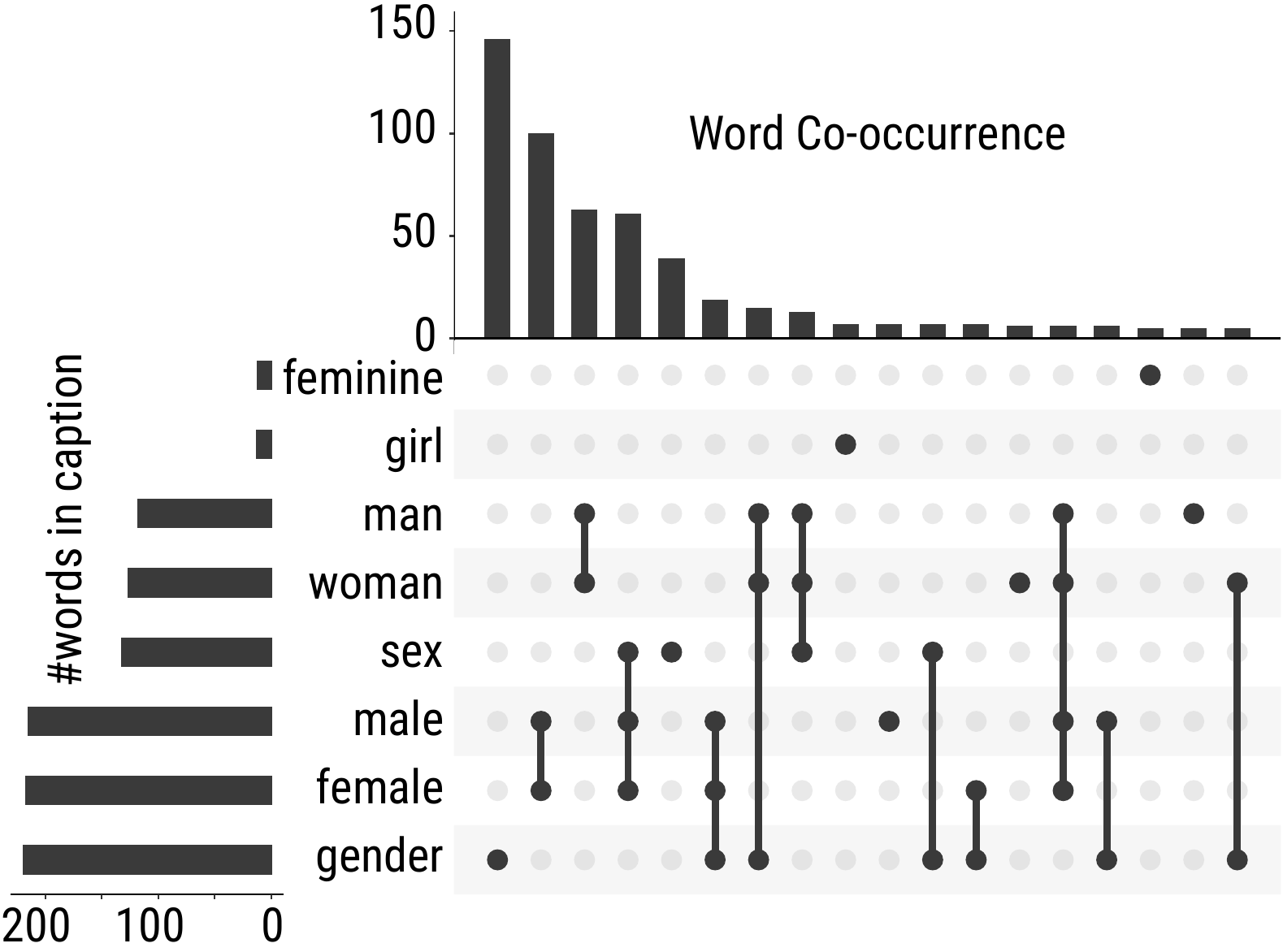}
 \caption{Association of gender-related words in figure captions. Minor associations ($<$ 5) have been cropped. }
 \label{fig:general_words_caption}
\end{figure} 

%

\subsubsection{General analysis}

\textbf{Captions.}
We summarize the occurrence of various keywords in captions in \cref{fig:general_words_caption}. The word \textit{gender} was the most frequent. It appeared in 38\% of all image captions. It was also used most often without any other words present. The association \textit{\{female, male\}} (in any particular order) was the second most frequent language. This set of words appeared in 97 captions, and with other words such as \textit{sex} (60\texttimes) or \textit{gender} (19\texttimes) in 196 captions (33.9\%). The word set \textit{\{women,men\}} was present in 18.2\% of all captions (62 alone and in 105 captions with other words). The fact that \textit{gender} most commonly appeared alone indicates that the visual encoding had to be explained in the chart itself, for example using labels or a legend, or that authors may have understood it as self-evident. 

The two words \textit{\{women, men\}} were almost as often mentioned with \textit{gender} (15\texttimes) and \textit{sex} (13\texttimes). As discussed in the background section and perhaps highlighted by these observations, the appropriate use of gender terminology is evolving and differs by field of study. 

\begin{figure}[tb]
 \centering
 \includegraphics[width=\columnwidth]{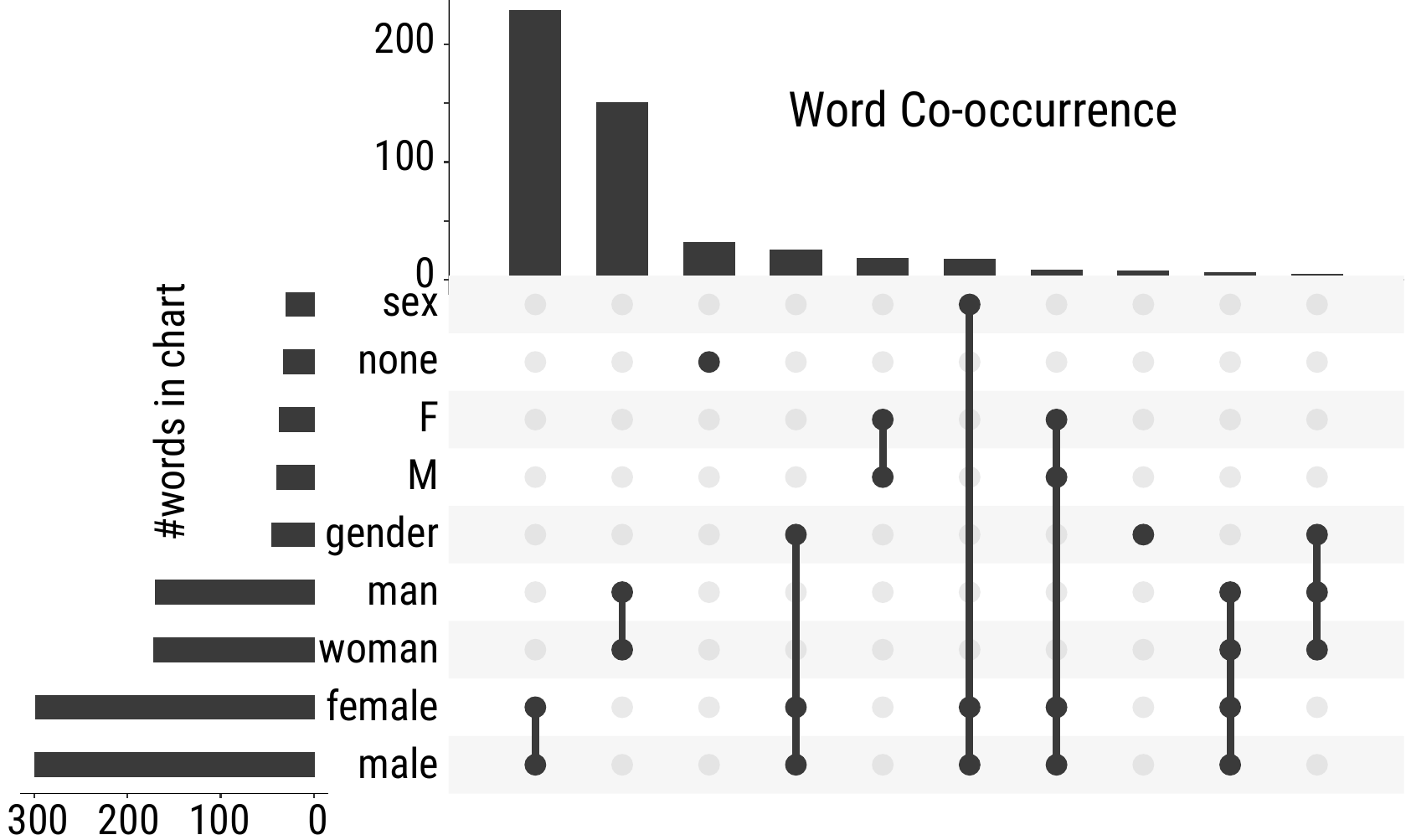}
  \vspace{-2em}
 \caption{Co-occurrence of gender-related words embedded in charts. Minor associations ($<$ 5) have been cropped.}
 \label{fig:general_words_charts}
  \vspace{-2.0em}
\end{figure}

\textbf{Text Embedded in Charts.} The words \textit{\{female,male\}} occurred most often together, in over half the images (51\%); by themselves in 217 figures, and in an additional 78 images with other words. The next most common co-occurrence was \textit{\{women,men\}}. It was used in 166 images (28.7\%), including 143 used alone.
Compared to the caption text, abbreviations were used more often in charts to describe gender (39 for \textit{M} and 37 for \textit{F}, compared to 2 for each in captions). The words \textit{gender} and \textit{sex} were used less in figure text (cf. \cref{fig:general_words_charts}).
We note that about half of our images came from the biology journal, so our data may over-represent the domain-specific use of words in biology. We next analyze the difference between biology and non-biology journals.

\subsubsection{Differences between biology and non-biology journals}\label{sec:wordsbyjournal}

As highlighted in the background section, gender-related words are used differently in different fields. Therefore, we study the differences between the biology journal and non-biology journals.

\textbf{Caption.} In non-biology journals, \textit{gender} alone was the most commonly used term (207\texttimes, 62\%). The second most common occurrence is \emph{\{female,male\}}. It appeared 31\texttimes\ alone and 61\texttimes\ associated with other words representing 18.3\% of captions. The set \textit{\{women,men\}} was almost as common and appeared 46\texttimes\ (13.6\%) including 28\texttimes\ alone.
In the biology journal's captions, the authors mainly used the words \textit{\{female,male\}} (66\texttimes\ alone, 135\texttimes\ associated with other words representing in total 55.6\% of all captions). The words \textit{\{women,men\}} were much less common. They appeared in 24.7\% of all captions; 34\texttimes\ alone and 60\texttimes\ associated with other words. 

\textbf{Text Embedded in Charts.}
Authors from non-biology journals used the pair \textit{\{female,male\}} in 48.8\% of images: 105\texttimes\ alone, 163\texttimes\ associated with other words. The pair \textit{\{women,men\}} was used in 30.8\% of images: 84\texttimes\ alone and 103\texttimes\ associated with other words. Moreover, the word \textit{sex} was almost never mentioned (only 12\texttimes\ in total).

Biology authors mainly used the terms \textit{\{female,male\}}: 112\texttimes\ alone, 132\texttimes\ with other words for in total 54.3\% of images. Similar to our finding for captions, the words \textit{\{women,men\}} were used less often for 25.9\% of images: 59\texttimes\ alone, 63\texttimes\ in total. 

\subsubsection{Analysis by period}
In order to understand if language use has evolved in the two time periods (2012--2017, 2018--2022), we analyzed the data separately. For captions, we did not see much of a difference in word use. The most frequent terms were the same for the two time periods: \emph{female, male, gender, sex, women, and men}. Their combinations also remained similar with \emph{\{female,male\}} being the most common pair in both periods. Similarly, we saw that the words \emph{female, male, women, men} were by far the most prevalent in both time periods. The word \emph{non-binary} was used five times in the more recent period and only once earlier. Still, the use of any nonconforming gender-related terms was very low. 

\subsubsection{Comparison of academic and public-facing images}
Some IIAB and TP images did not come with descriptive text that could be identified as a caption. Of the 96 public-facing images, 32 used gender terms either in their title or in the description of the visualization. The preferred association was \textit{gender} alone (8) followed by \textit{\{women,men\}} (6) and \textit{\{female-male\}} (4).
Just over half the images (52\%) used the words \textit{\{female,male\}} inside their charts: 50\texttimes\ in total, including 17\texttimes\ alone and 15\texttimes\ paired with the word \textit{gender}. The pair \textit{\{women,men\}} were used in 33\% of images (32\texttimes, including 14\texttimes\ alone). This result is consistent with our text analysis in scientific journals.

\subsubsection{Summary}
In both captions and charts, the words \textit{\{female,male\}} were the most used to refer to gender or sex. Non-biology journal authors tended to use the word \textit{gender} by itself in captions (rather than \textit{sex}) coupled with the use of \textit{\{female,male\}} to refer to gender represented in a chart. 

\subsection{Nonconforming and not reported gender}\label{sec:nonbinaryvis}

We were able to extract images that included nonconforming or ``other'' gender from our coding of color and shape used in the figures as well as from the use of words in captions and image text. In total, we found only ten images from researchers that represented data for individuals whose gender was either intersex (1), non-binary (5), transgender (3), or unknown/not reported (1). Nine of these images comprised a single nonconforming gender. The tenth image represented three different kinds of intersex people. Nine images represented nonconforming gender compared to both women and men and the last one compared cisgender women to transgender men.
Of the 96 relevant images from the public dataset, we found seven images (four from IIAB and three from TP) representing another gender than women and men. Four images represented an unknown or undefined gender, one represented non-binary people (compared to women and ``others''), and the last two encoded transgender and non-binary people. 

\subsubsection{Encodings}

The authors of nonconforming gender scientists' images encoded gender almost exclusively by using color (5) or position (4). Two of them used both color and position, while one used a table to display their data. When colors were included in the visualizations, three color combinations which differ from red to represent women and blue to represent men, were proposed. One of these images represented women, men, and non-binary, respectively, in yellow, blue, and pink colors. Two other images represented women in orange and men in green; while ``unknown'' gender was blue for one and non-binary was grey for the other. One visualization used colors falling between red and blue for intersex individuals, while the last image used white for cisgender women and grey for transgender men. 

As with academic visualizations of nonconforming gender people, designers of public-facing images exclusively used colors to distinguish genders. Two images from TP used the Tableau standard color palette, and the last one used a color gradient from dark blue for ``male'' to light green blue for non-binary and transgender people. The four images from IIAB compared data from individuals with ``unknown'' or ``unspecified'' gender (3) or transgender or non-binary individuals (1). Two representations used an ungendered category encoded in orange or grey, while women's data was coded in pink and men's in blue. The last image, including an ``unknown'' gender represented this group in green. Finally, one image used brown-red for trans and non-binary people.

\subsubsection{Text analysis}

Images from researchers and practitioners that represent nonconforming gender seem to use a wider range of terms. Unsurprisingly, the term \textit{non-binary} appeared ten times (in eight charts and two captions), and \textit{transgender} was used five times to describe gender (in two captions and three legends). 
In addition, the use of \textit{sex} to describe gender almost disappeared (only two occurrences among the 17 images).
However, the use of \textit{\{women,men\}} and \textit{\{female,male\}} in the caption was still dominant (8\texttimes\ vs. 7\texttimes). In charts, \textit{female} and \textit{male} terms were more often used (9) than \textit{women} and \textit{men}. We noticed that eight uses (among nine) of \textit{female} and \textit{male} in charts came from TP images.

\subsection{Discussion and limitations of our analysis}
Several different reasons may have led to our findings. Stereotypes might be a quick conclusion to arrive at, but we found little evidence that, for example, the famous blue/pink color association was frequently used. Another possible influence is the choice of tools authors used to create graphics for scientific papers, such as R (ggplot), Tableau, or Excel. These tools propose a color palette by default. For instance, Tableau, ggplot (R), and Excel use reds, greens, and blues as their first categorical colors. Ggplot uses squares and circles as the first two shapes. Excel uses diamonds and squares. Yet, each of these tools offers ways to customize visualizations. Sometimes customizations might have been made because of domain-specific conventions, as discussed for Kinship or Pedigree diagrams, but we cannot trace customization choices back without consulting the original authors. 

Another limitation of our work is the selection of figures based on a set of 22 words in the text of captions. We might not have captured graphics that used abbreviated genders or used icons in captions. Images about specific jobs (\eg actors vs.\ actresses) would also have been missed. In addition, we only selected a subset of journals that we found representative and our choice of public images was limited to just two sources. Yet, it is impossible to provide a complete analysis of the entire research and public corpus of visualization images. We are confident that our choices provide a good basis for discussion, particularly since results were quite similar across our sources.

As we mentioned before, the few representations of nonconforming gender visualizations prevent us from producing an in-depth analysis. We searched for journals potentially relevant to our analysis, such as the International Journal of Transgender Health. Unfortunately, these journals did not provide enough images to be analyzed. This may be remedied in the near future with a focused study based on a very broad selection of sources.

\section{Considerations for Visual Gender Representation} \label{sec:considerations}

By creating visualizations about gender data, designers and researchers have the unique opportunity to represent marginalized populations and support discussions on gender differences. In the visualization and HCI communities, we often study and represent data about humans while at the same time discussing data in abstract terms \cite{Morais:2022:Anthropographics}. We should not underestimate visual representations' impact on people’s sense of belonging. Impactful gender visualizations require careful consideration of how gender is encoded. Our analysis aimed to study how gender data is currently visualized as a basis for constructive conversation. We purposefully avoid talking about guidelines for gender representation, as the conversation about gender representation is ongoing, and guidelines may quickly lead to stereotypes or conflict with guidelines for other demographics, such as race \cite{Schwabish:2021:Harm}. We highlight some important aspects of gender visualizations below and hope readers will take away key points to consider when creating their own gender data visualizations.

\subsection{Inclusive data collection matters}

Data visualizations necessarily need data. What a visualization can show is inherently tied to the underlying data. Yet, when a visualization designer is faced with incomplete data, for example, data with only two genders, one option is to acknowledge the absence of broader gender definitions. This acknowledgment may be visual, presented in the caption, or embedded in the charts. Tovanich et al.\ \cite{tovanich:2022:gender} in their analysis of the gender representation at IEEE VIS, for example, mention in the introduction that `` [they] \textit{are not able to capture the non-binary and fluid nature of gender and acknowledge this shortcoming}''. 

Visualization designers sometimes have the opportunity to capture data, for example, in user studies. The construction of survey instruments on demographic information needs to be carefully thought out and has been extensively studied in many domains \cite{Scheuerman:2021,Bauer:transgender:2017,Colaco:2021:Collection,Chuck:2013:Collection,westbrook2015new,Brown:2020:Collection}. Recently, the US federal government released a 15-page document promoting best practices for collecting gendered data \cite{USA:2023:Guidelines}. In these guidelines, the authors emphasize the importance of having a write-in answer option and provide six different ways to ask about gender. Scheuerman et al.\ \cite{Scheuerman:2021} analyzed how binary and non-binary individuals perceive gender input forms. The authors provided concrete suggestions on what not to do but also offered design considerations such as abandoning the terms \textit{female}, \textit{male}, and \textit{sex}. Recently, Beischel et al. \cite{Beischel:2021:Collection} proposed a visual tool for collecting and visualizing participants' gender and sex. Using this tool, participants represent themselves on a circular diagram that includes different dimensions of gender or sex.

In addition to creating inclusive data collection instruments, it is important that researchers implement data protection measures and minimize risks for participants. In this context, we also recognize that some cultures and social groups do not acknowledge nonconforming gender. Therefore, the collection and visualization may not be appropriate.

\subsection{Increasing awareness and reflection}

Data visualization is often a compromise between legibility, aesthetics, and usefulness. To manage these trade-offs, designers must make decisions that can be guided by their culture, the message they want to convey, and other factors. Gender data visualizations are no exception. 

Stereotypes and biases are difficult to address because they often go unnoticed. Discussing our practices as individuals and as a community is necessary to surface stereotypes and biases. For example, at the beginning of this project, we were well aware of existing discussions around ``pink is for girls and blue is for boys,'' but it was not clear to us how limited the color palette for men was and how prevalent red was as a color to represent women.
 
Another possible bias is that researchers might think of gender data as something irrelevant or periphery. We collected more than 30k images from CHI papers and 14k images from VIS, and only 0.2\% reported gender visually or in a caption. It is crucial to remedy this lack of material: doing so might support the study of gender differences and allow the visualization community to build a foundation for meeting the present and future needs for equality and inclusiveness. 

Finally, a possible explanation for our results relates to the authors and image creators. Studies from different scientific domains have shown an overrepresentation of men in research\cite{tovanich:2022:gender,Holman:2018:gender,Lariviere:2013:gender}. If the creators of our analyzed figures are mostly men, they may have unconsciously selected colors they associate with themselves or their characteristics. Unfortunately, we do not have the necessary data to study whether men create different graphics than women. Yet, considering how one's own gender might influence the visualization encoding types we use is worth some thought and even future empirical work.

\subsection{Thoughtful design is critical}
We now discuss data encodings in more detail. Designing a visualization is not just a matter of assembling and mapping elements to visual variables. The use of color, shape, or texture conveys a message that the designer can purposefully influence. In this section, we will go through multiple visual attributes to discuss related design considerations.

\subsubsection{Gender-stereotyped colors}

Colors are not neutral and often have connotations. Blue is usually associated with positive valence, positive emotions, and masculine attributes \cite{Jonauskaite:2019}. Several studies show that gender is also associated with color brightness \cite{Mochizuki:brightness:2022,semin:brightness:2014}. We found it particularly interesting that women were more often displayed in brighter tones when the images used achromatic colors; a finding that supports prior work \cite{semin:brightness:2014}. These color associations recall images of brides in white dresses and men in black suits. While it is perhaps impossible to foresee or even to account for all possible stereotypical associations when assigning colors, shapes, or textures to encode the gender variable, designers should minimally consider if any obvious associations come to mind. 

Some professional visualization designers have tried to move away from traditional gendered colors and proposed different color encodings. In her blog, Muth \cite{Muth:2018:alternative} reports examples that use purple, green, or orange as the key colors. These colors started to appear in the visualization community (\eg \cite{tovanich:2022:gender}) but remain uncommon (per \cref{fig:sankey_color}). Another possible solution, at the risk of a less vivid aesthetic, may be to remove color as a distinguishing factor altogether. For example, labels are sometimes enough to identify a gender when the marks for each gender are clearly separated in space, such as in a grouped bar or split chart. However, removing color becomes harder or impossible in other types of visualizations, such as streamgraphs or heatmaps.

Semantically resonant colors are those people associate with certain concepts\cite{lin:resonant:2013}. For example, for data about fruits, bananas could be shown in yellow and cherries in red. Prior work has shown that choosing semantically resonant colors improves people's performance in certain data reading tasks \cite{lin:resonant:2013}. Further discussions are needed around gender and semantically resonant colors. For example, does moving away from the conventional red for women create ``ineffective'' visualizations? Is the risk of perpetuating color stereotypes with semantically resonant gender colors worth creating charts that people might read more easily? Prior work showed that inverting gendered associations (presenting women's first names in dark color and men's first names in bright) had a negative impact. The participants associated the first names with gender slower when stereotypes were inverted \cite{semin:brightness:2014}, but more work is needed to understand the real effect of inverting gendered associations on visualization understanding. Further, what is the effect of stereotypical colors on higher-level tasks related to decision-making? The research space is wide open, and we hope others will engage and contribute to help answer some of these questions.

\subsubsection{Reflection on other encodings} 
Following color, the other most commonly used encodings were separated charts, positions, line patterns, and shapes. Stroessner et al. \cite{Stroessner:shape:2020} found that abstract shapes were also gendered in self-perception and implicit associations tasks. They found that circles represent women's attributes while squares represent men's attributes, which matches the existing conventions we mentioned in \cref{sec:shape}. Our analysis did not support these results, as circles and squares were almost equally common for both genders. Moreover, shapes were often combined with other encodings, such as colors. Expanding the work on semantically resonant colors to shapes might be worthwhile. 
As for colors, a key consideration is that shapes may have different connotations \cite{Pinna:2011:Shapes}.

Arranging elements in visualization is a crucial design choice \cite{munznerVisualizationAnalysisDesign2014} that can play a role in interpreting entire charts, marks, or words in a chart and thus have an impact on how people read gender data. 
On the level of entire charts, for example, panels use spatial location to separate information into separate charts, meaning that one gender may come first or on top of the other (see \cref{fig:teaser} (C)). Similarly, in a grouped bar chart, the order of marks means that one gender will likely be read first (see \cref{fig:teaser} (B)). Even legends include an order that can be specifically chosen (see \cref{fig:teaser} (D) and (E)). The order (\textit{male-female} versus \textit{female-male}) may imply a hierarchy between groups, and its impact deserves to be studied in the future.

Regrettably, we did not code the arrangement of gender representation in our study. Yet, during the coding, we began asking ourselves whether men were more frequently displayed on the left or top. In particular, when the words used in the charts were \textit{\{female,male\}}, it appears that the legends were often not alphabetical. We have no information about whether visualization designers chose this positioning or whether it was an automatic result of the data's order or frequency. Further, we cannot say whether the order even matters to people's perception of the genders represented, which would merit further study. 

Line patterns and textures, both of which repeat the same sets of elements (dashes and shapes, respectively), have not been extensively studied. As far as we know, no study links gender to these encodings. However, line patterns were extensively used in line charts, especially in greyscale ones. Further investigations are required to assess to what extent textures and line patterns can be gender-related.

\begin{figure}[tb]\setlength{\belowcaptionskip}{-8pt}
\centering
    \includegraphics[width=.62\columnwidth]{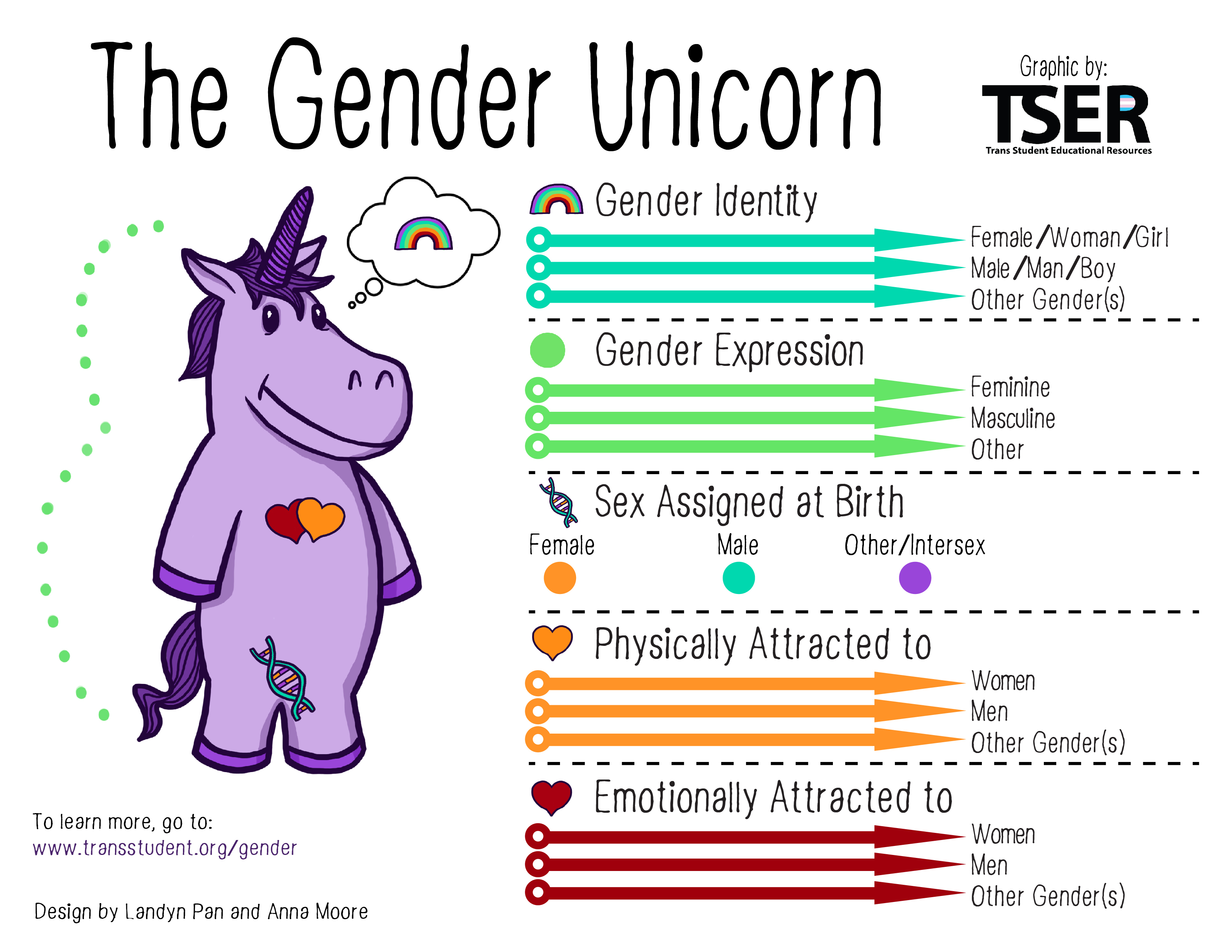}
    \includegraphics[width=.37\columnwidth]{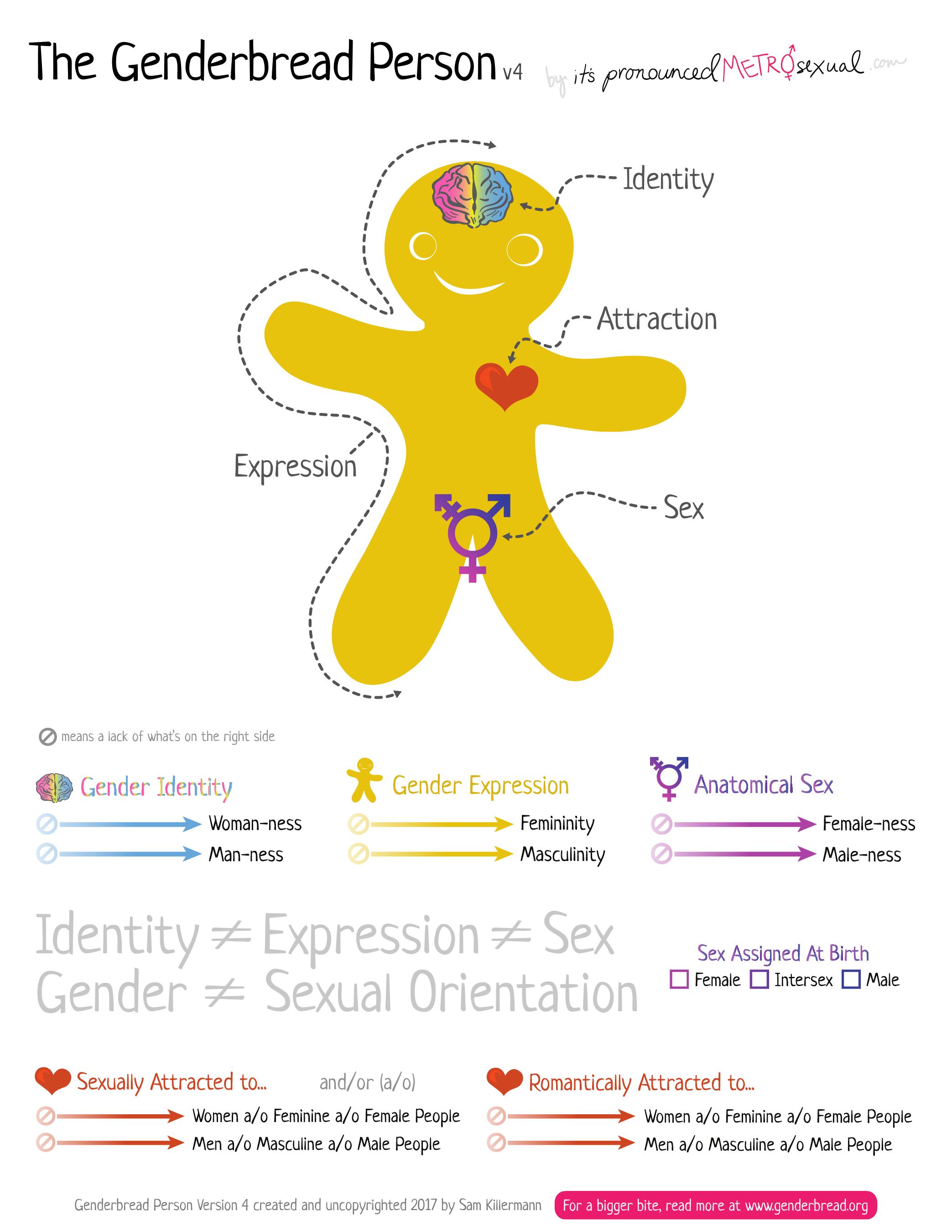}
  \vspace{-2em}
    \caption{The Gender Unicorn \cite{TSER:2015:Unicorn} and the Genderbread Person \cite{GBP:2018:GBP}. Image Permissions: (left) Creative Commons and (right) Uncopyrighted}

    \label{fig:gender_tools}
  \vspace{-1em}
\end{figure}


\subsection{Choose your language carefully}
The pair \textit{\{female,male\}} was the most common in our images. However, the words \textit{\{women,men\}} were also common, and notably, they have become more common in recent years in the biology community. Our text analysis confirms that in non-biology journals, the most common practice is also to refer to participants as \textit{female} and \textit{male}, which is consistent with previous work in HCI \cite{Offenwanger:2021}.

Gendered language, including specific words, should be chosen carefully. Yet, for international researchers  from different countries, cultures, and different native languages, staying up-to-date 
can be challenging. As explained in the background section, the language around gender identities is still evolving. Some tools can help researchers and citizens use more inclusive language. For instance, Trans Student Educational Resources proposed the Gender Unicorn \cite{TSER:2015:Unicorn} (\cref{fig:gender_tools} --- left), an infographic that sums up the different dimensions of gender. The Genderbread Person is another engaging resource that explains the basic concepts of gender and sex (\cref{fig:gender_tools} --- right).

In the HCI community, some researchers have started to discuss the perception of conventional language used to talk about study participants (\eg users or participants) \cite{Bradley:2015:Gendered} and the use of terms, such as \textit{sex}, \textit{female} or \textit{male} or traditional binary barriers \cite{Scheuerman:2021,Offenwanger:2021}. Especially in data collection, asking about binary genders is perceived as limited, with a feeling of unease about choosing between only two genders by both binary and non-binary participants\cite{Scheuerman:2021}. 

\subsection{More inclusive representation}

\begin{figure}[tb]
 \centering 
 \includegraphics[width=0.95\columnwidth]{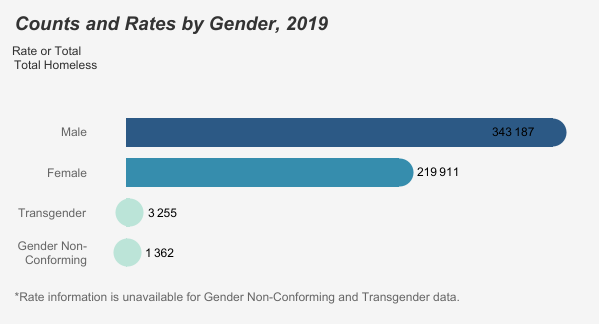}
  \vspace{-1em}
 \caption{Data visualization using different scales to highlight transgender and gender-nonconforming data \cite{janosko_2020}.}
 \label{fig:non-binary-scale}
  \vspace{-2.0 em}
\end{figure}

We found very few visualizations in our corpus that included nonconforming genders. Assuming we have, however, followed good practices and collected such data, how should we represent it? Most likely, the small size of gender-nonconforming people’s data will be dwarfed by data collected on women and men. It is worth looking at dedicated visualization techniques for addressing data series of different magnitudes. For example, one could break a chart apart and plot the smaller numbers below, next to a break with a dedicated axis, or use log scales \cite{isenberg:2011:dualscale}.
Another solution could be switching to a completely different representation mechanism for gender nonconforming data, such as anthropographics. Ultimately, the solution will depend on the visualization type chosen for the problem. We saw one clever representation in the TP corpus that used a different scale (cf. \cref{fig:non-binary-scale}). 

An interesting dilemma is posed here regarding our ethical obligations not to falsify, amplify, omit, or change data in visualizations. To what extent should we ensure the visibility of minority groups when it requires more space than the data values warrant? We hope that researchers will continue to think about this challenge and come up with replicable ways to represent more than two genders.

\section{Discussion}
The importance of inclusive visualization is not limited to gender alone. Some personal attributes, including age, ethnicity, and disability, in addition to gender, have received the status of ``protected attributes'' in some countries, including the UK \cite{UK:2010:EqualityAct} and the US \cite{USA:CivilRight}. The purpose of this action is to safeguard against discrimination based on these attributes. For example, the use of protected attributes in AI is strictly regulated \cite{Colmenarejo:2022:EUAI}. These attributes deserve careful and thorough attention from the visualization community, and others have already started the discussion. Dhawka et al. \cite{Dhawka:2023:Race} have made an especially significant contribution by proposing a set of challenges and opportunities surrounding the visualization of race-related diversity. We hope their work and ours will inspire researchers to continue discussions of how, when, and whether to visualize protected attributes. These discussions need to start before we can consider how recent AI developments might aid us in generating and analyzing visualizations of protected attributes.

Focusing on gender, there are several interesting questions to discuss and explore further. Comparing a dominant and a marginalized group can result in discrimination against the marginalized group \cite{Cecchini:2019:Stereotypes}. For example, by emphasizing data about transgender or gender nonconforming individuals (see \cref{fig:non-binary-scale}), we might invite stereotyped reasoning \cite{Cecchini:2019:Stereotypes}. Cecchini \cite{Cecchini:2019:Stereotypes} proposed three guiding principles for facing this dilemma: raising awareness about ethics, considering ethics throughout one's work, and empowering the represented people regarding their opinions and answers. In line with these principles, every study of protected attributes must carefully consider ethical risks.

\section{Conclusion}
The goal of our paper is to raise awareness and start a discussion on how to represent gender in visualizations. We have pointed to many open challenges and questions that the wider scientific community has yet to address (see \cref{sec:considerations}). Our work is grounded in the analysis of both scientific and public-facing visual gender representation. Our analysis focuses on the different ways scientists distinguish gender visually in images and what language they use to describe and distinguish gender. 
Future discussion around considerations of gender visualization should aim to reach more inclusive data visualization practices. Such carefully designed visualizations may be a first step towards more inclusive data representations in general.
We hope that our work will be a starting point for deeper reflection both in terms of visualization and in terms of representing gender and people. 

\section*{Supplemental Materials}
\label{sec:supplemental_materials}
All supplemental materials are available on OSF at \href{https://osf.io/v9ams/}{osf.io/v9am}, released under the \href{https://creativecommons.org/licenses/by/4.0/}{CC BY 4.0}.
In particular, they include (1) Excel files containing the data used for the analysis, (2) figures from the paper (3) the screen capture of our tool mentioned in \cref{sec:coding}, (4) the R scripts used to compute results and create \cref{fig:encodings_general,fig:colors_per_year,fig:colors_per_journal,fig:colors_public_facing,fig:general_words_caption,fig:general_words_charts} 
, (5) HTML and  code used for creating \cref{fig:sankey_color} and \cref{fig:shapes_gender}, (6) the user guide given to the coders and (7) a demonstration video of the coding tool.

\section*{Figure Credits}\label{sec:figure_credits}
With the exception of those images from external authors whose licenses/copyrights we have specified in the respective figure captions, we as authors state that all of our own figures, graphs, plots, and data
tables in this article (\ie those not marked) are and remain under our own personal copyright, with the permission to be used here. We also make them available under the \href{https://creativecommons.org/licenses/by/4.0/}{Creative Commons Attribution 4.0 Inter-
national (CC BY 4.0)} license and share them at \href{https://osf.io/v9ams/}{osf.io/v9am}.

\vspace{-0.1em}

\acknowledgments{%
The authors wish to thank Chaitanya Pohnerkar and Praharsh Deep Singh for their help during the coding phase, Jian Chen for the figures extraction from PDF, Rui Li for sharing the skeleton of the online coding tool, and Stephanie Manuzak for proof-reading the first version of the paper.
 This work was supported in part by the Action Exploratoire (AeX) EquityAnalytics from Inria.%
}
\vspace{-0.1em}
\bibliographystyle{abbrv-doi-hyperref-narrow}

\bibliography{template}
\end{document}